\documentclass[aps,prl,twocolumn,superscriptaddress]{revtex4-2}

\usepackage{empheq}
\usepackage{graphicx}
\usepackage{epsfig}
\usepackage{amsmath,amsthm}
\usepackage{color}
\usepackage{verbatim}
\usepackage{ulem}
\usepackage{siunitx}
\usepackage{lineno}

\begin{document}

\preprint{unpublished manuscript}


\title{Engineering altermagnetic symmetry to enable anomalous Hall response in Cr$_{1-x}$Mn$_x$Sb}

\author{Miriam G. Fischer$^\dagger$}
	\affiliation{Institut f\"ur Physik, Johannes Gutenberg-Universit\"at, Staudinger Weg 7, 55128 Mainz, Germany}
\author{Lukas Odenbreit$^\dagger$}
	\affiliation{Institut f\"ur Physik, Johannes Gutenberg-Universit\"at, Staudinger Weg 7, 55128 Mainz, Germany}
\author{Olena Gomonay$^\dagger$}
	\affiliation{Institut f\"ur Physik, Johannes Gutenberg-Universit\"at, Staudinger Weg 7, 55128 Mainz, Germany}
\author{Sonka Reimers}
	\affiliation{Institut f\"ur Physik, Johannes Gutenberg-Universit\"at, Staudinger Weg 7, 55128 Mainz, Germany}
\author{Pascal Manuel}
	\affiliation{ISIS Neutron Facility, STFC Rutherford Appleton Laboratory, Chilton, Oxfordshire, OX11 0QX, UK}
\author{Dmitry Khalyavin}
	\affiliation{ISIS Neutron Facility, STFC Rutherford Appleton Laboratory, Chilton, Oxfordshire, OX11 0QX, UK}
\author{Sean Langridge}
	\affiliation{ISIS Neutron Facility, STFC Rutherford Appleton Laboratory, Chilton, Oxfordshire, OX11 0QX, UK}
\author{Jairo Sinova}
	\affiliation{Institut f\"ur Physik, Johannes Gutenberg-Universit\"at, Staudinger Weg 7, 55128 Mainz, Germany}
\author{Thibaud Denneulin}
	\affiliation{Ernst Ruska-Centre for Microscopy and Spectroscopy with Electrons, Forschungszentrum Jülich, 52425, J\"ulich, Germany}
\author{Joseph V. Vas}
	\affiliation{Ernst Ruska-Centre for Microscopy and Spectroscopy with Electrons, Forschungszentrum Jülich, 52425, J\"ulich, Germany}
\author{Rafal E. Dunin-Borkowski}
		\affiliation{Ernst Ruska-Centre for Microscopy and Spectroscopy with Electrons, Forschungszentrum Jülich, 52425, J\"ulich, Germany}
\author{Mathias Kl\"aui}
	\affiliation{Institut f\"ur Physik, Johannes Gutenberg-Universit\"at, Staudinger Weg 7, 55128 Mainz, Germany}
\author{Martin Jourdan}
	\affiliation{Institut f\"ur Physik, Johannes Gutenberg-Universit\"at, Staudinger Weg 7, 55128 Mainz, Germany}
			\email{Jourdan@uni-mainz.de}

\begin{abstract}
Altermagnets are a promising class of materials for spintronic applications. However, compounds that simultaneously combine the symmetry required to support an anomalous Hall effect with good metallic conductivity and magnetic ordering temperatures well above room temperature have remained elusive. Here, we demonstrate that partial substitution of Cr by Mn in epitaxial CrSb(100) thin films provides a viable route to engineer the combined structural and magnetic symmetry necessary to enable an otherwise symmetry-forbidden anomalous Hall effect. We systematically explore the magnetic phase diagram of Cr$_{1-x}$Mn$_x$Sb thin films including neutron diffraction. This allows us to identify a magnetic symmetry that supports the anomalous Hall effect, which we demonstrate in Cr$_{0.75}$Mn$_{0.25}$Sb. Guided by Landau theory, we model the field-driven reorientation of the N\'eel vector and the resulting anomalous Hall response, achieving good qualitative agreement with the experimental observations.   
\end{abstract}



\maketitle

* email: Jourdan@uni-mainz.de\\
$^\dagger$ These authors contributed equally to this work.

\section{Introduction}
Altermagnets are promising for spintronics due to their potential to combine key advantages of both conventional antiferromagnets—such as ultrafast spin dynamics—and ferromagnets, namely, the ability to generate spin-polarized currents. 
Altermagnetism arises from the interplay between a specific crystal symmetry and a compensated collinear arrangement of identical magnetic moments. To realize altermagnetism, the lattice sites of two ferromagnetically ordered sublattices—whose moments are aligned antiparallel to each other—must be related by a combined translational and rotational symmetry operation. This symmetry leads to a direction-dependent exchange interaction, manifesting as an alternating spin splitting of the electronic bands \cite{Sme19,Hay19,Maz21,Sme22a,Sme22b}. Furthermore, despite zero net magnetization, an anomalous Hall-effect (AHE) enabling the read-out of the direction of the N\'eel vector can occur \cite{Sme19}. In contrast to the altermagnetic splitting of the electronic bands, this effect requires spin-orbit coupling. The AHE always appears together with weak ferromagnetic behaviour, i.\,e.\,canted spin moments due to a Dzyaloshinsky-Moriya type interaction \cite{Dzy58}, as they share the same symmetry based requirements to exist \cite{Gon23}.

The most compelling experimental evidence for altermagnetic band splitting has been provided by angle-resolved photoemission spectroscopy (ARPES) studies of the isostructural intermetallic compounds MnTe \cite{Kre24, Osu24, Lee24} and CrSb \cite{Rei24, Din24}, classified as g-wave altermagnets due to their hexagonal symmetry. The magnetically ordered state of both compounds consists of ferromagnetic (001) planes, which are coupled antiferromagnetically along the c-axis. However, in MnTe, the N\'eel vector lies within the hexagonal c-plane and aligns along the [1$\bar{1}$0] crystallographic direction enabling an AHE \cite{Gon23}. In contrast, in CrSb the N\'eel vector points along the crystallographic c-axis \cite{Sno52}, a direction that does not provide the magnetic-space symmetry required for an AHE \cite{Zho25}.

This is disadvantageous for potential spintronics applications, as otherwise CrSb stands out due to its relatively high magnetic ordering temperature of 700~K \cite{Tak63} and its metallic band structure with a significant altermagnetic spin splitting ($\simeq0.6$~eV) positioned just below the Fermi energy \cite{Rei24}. Thus, a proper combination of the magnetic and electronic properties of CrSb and MnTe would result in a very promising altermagnet. In epitaxial CrSb(100) thin films, this requires tilting the magnetic easy axis away from the crystallographic c-axis.

Here, we demonstrate that partial substitution of Cr with Mn in the parent compound CrSb enables tuning of the magnetocrystalline anisotropy and engineers the symmetry and magnetic order to allow for an AHE signal. While the crystallographic symmetry is deliberately maintained, the modification of the chemical composition enables tailoring the easy directions of the antiparallel coupled magnetic moments. For the composition Cr$_{0.75}$Mn$_{0.25}$Sb, we observe an AHE in (100)-oriented epitaxial thin films, which is fully consistent with the expected altermagnetic properties associated with the magnetic space symmetry of this specific sample composition. We explain the origin of the magnetic field dependence of the AHE and its relation to the orientation of the N\'eel vector by a Landau-type phenomenological model based on symmetry considerations.

\section{Results and discussion}

{\bf Sampling the magnetic phase diagram of Cr$_{1-x}$Mn$_{x}$Sb}

\begin{figure*}
\includegraphics[width=2.0\columnwidth]{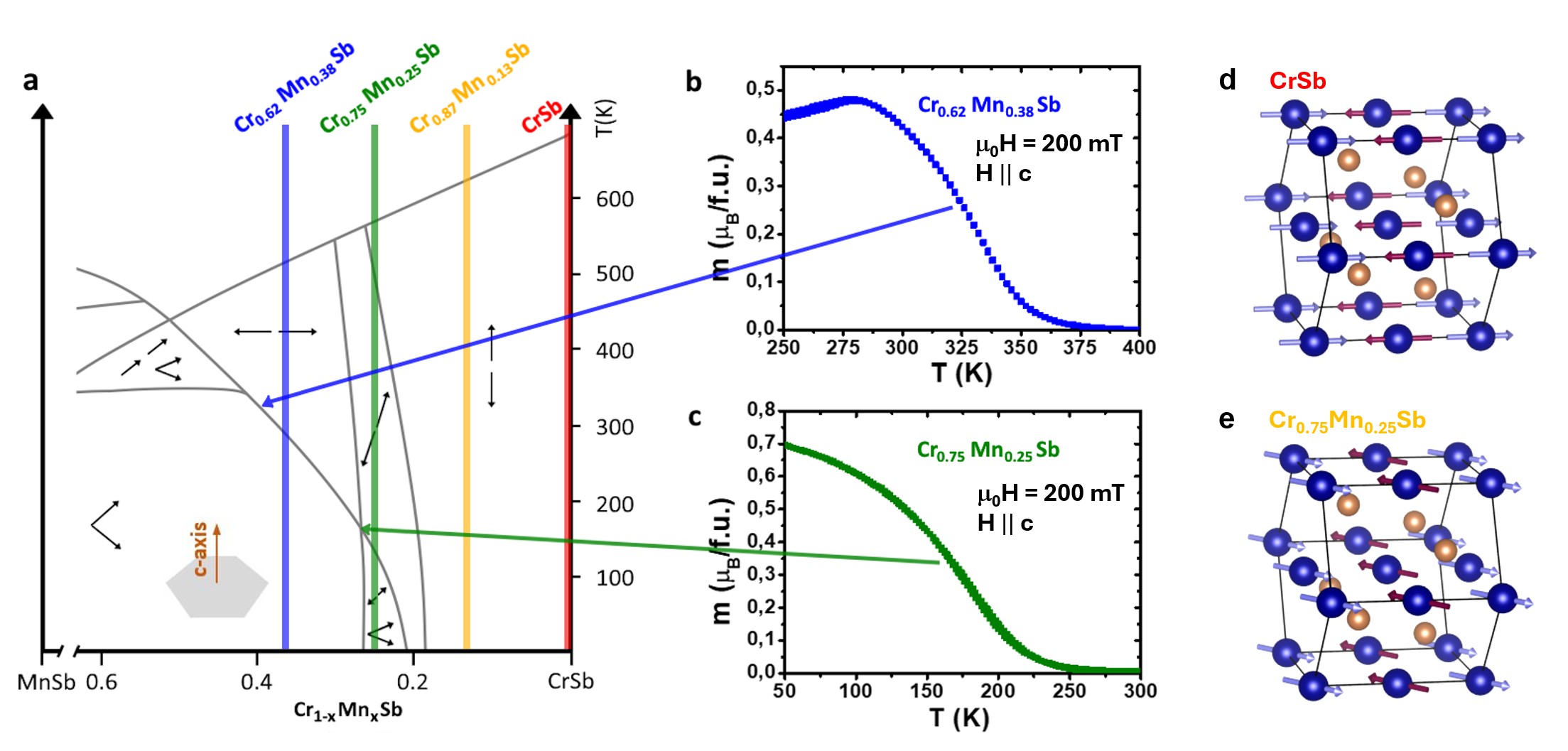} 
\caption{\label{Fig1} 
{{\bf Magnetic order of Cr$_{1-x}$Mn$_{x}$Sb samples. a} Magnetic phase diagram of Cr$_{1-x}$Mn$_{x}$Sb bulk samples adapted from \cite{Rei82} with color bars indicating the different compositions sampled by our epitaxial thin films. The black arrows represent the direction of the magnetic moments based on neutron diffraction of bulk single crystals \cite{Rei82}, where upwards and downwards align with the crystallographic c-axis and left and right correspond to moments within the haxagonal plane. 
{\bf b} Temperature dependent magnetization indicating the phase transition of a Cr$_{0.62}$Mn$_{0.38}$Sb(100) thin film (sweep rate 1~K/min). 
{\bf c} Temperature dependent magnetization indicating the phase transition of a Cr$_{0.75}$Mn$_{0.25}$Sb(100) thin film (sweep rate 1~K/min). Thin films with lower Mn content do not exhibit such a transition. Panels {\bf d} and {\bf e} illustrate the crystal structure of Cr$_{1-x}$Mn$_{x}$Sb with magnetic moment configurations corresponding to the parent compound CrSb ({\bf d}) and to Cr$_{0.75}$Mn$_{0.25}$Sb ({\bf e}). In the latter case, the easy axis is tilted away from the crystallographic c-axis, enabling an AHE. The pink plane corresponds to the substrate plane.} 
} 
\end{figure*}

Motivated by an early report of a complex magnetic phase diagram in bulk Cr$_{1-x}$Mn$_{x}$Sb samples \cite{Rei82}, we grew corresponding (100)-oriented epitaxial thin films on GaAs(110) substrates by DC sputtering (see {\it Experimental Methods} and {\it Supplementary Information}). By varying the Cr-to-Mn ratio, we accessed distinct magnetic phases, whose accurate identification is essential in the context of altermagnetism. Based on their composition and magnetic properties, we assessed whether the thin films can be assigned to the corresponding regions of the bulk phase diagram reported in Ref.\,\cite{Rei82}, which includes  both compensated magnetic as well as ferromagnetic phases. These magnetic phases are represented by the black arrows in Fig.\,1, panel {\bf a} and were previously identified by neutron diffraction studies on bulk single crystals. In the present work, we focus on the room temperature compensated phase with the magnetic easy axis tilted away from the hexagonal c-axis, corresponding to the composition Cr$_{0.75}$Mn$_{0.25}$Sb. This phase is of particular interest because it is the only one expected to exhibit an anomalous Hall effect in our measurement geometry. 

In Fig.\,1, we classify the epitaxial thin film samples within the magnetic phase diagram based on their composition, ferromagnetic transition temperature, and magnetic anisotropy. 
Panel {\bf a} shows the relevant region of the bulk phase diagram {\cite{Rei82}} with the compositions of three epitaxial Cr$_{1-x}$Mn$_{x}$Sb(100) thin films as determined by energy dispersive X-ray spectroscopy (EDX) performed in a transmission electron microscope (TEM). Panels {\bf b} and {\bf c} display temperature dependent magnetization measurements of the Cr$_{0.62}$Mn$_{0.38}$Sb and Cr$_{0.75}$Mn$_{0.25}$Sb films, respectively, aquired in a magnetic field of 200~mT. Both curves exhibit a transition from compensated order to a ferromagnetic state at temperatures consistent with the bulk phase diagram shown in panel {\bf a}.
In contrast, and again in agreement with the bulk phase diagram, such a temperature driven transition is absent for CrSb and Cr$_{0.87}$Mn$_{0.13}$Sb thin films.

The magnetic properties of the four epitaxial thin-film compositions investigated here are consistent with the magnetic phase diagram of Cr$_{1-x}$Mn$_x$Sb single crystals (Fig.~1a) reported in Ref.~\cite{Rei82}, indicating that the magnetic structures of the thin films are comparable to those of the bulk material of the same composition. To directly probe the magnetic order, we next employ neutron diffraction.
  
\vspace{5pt}

\begin{figure*}
\includegraphics[width=1.65\columnwidth]{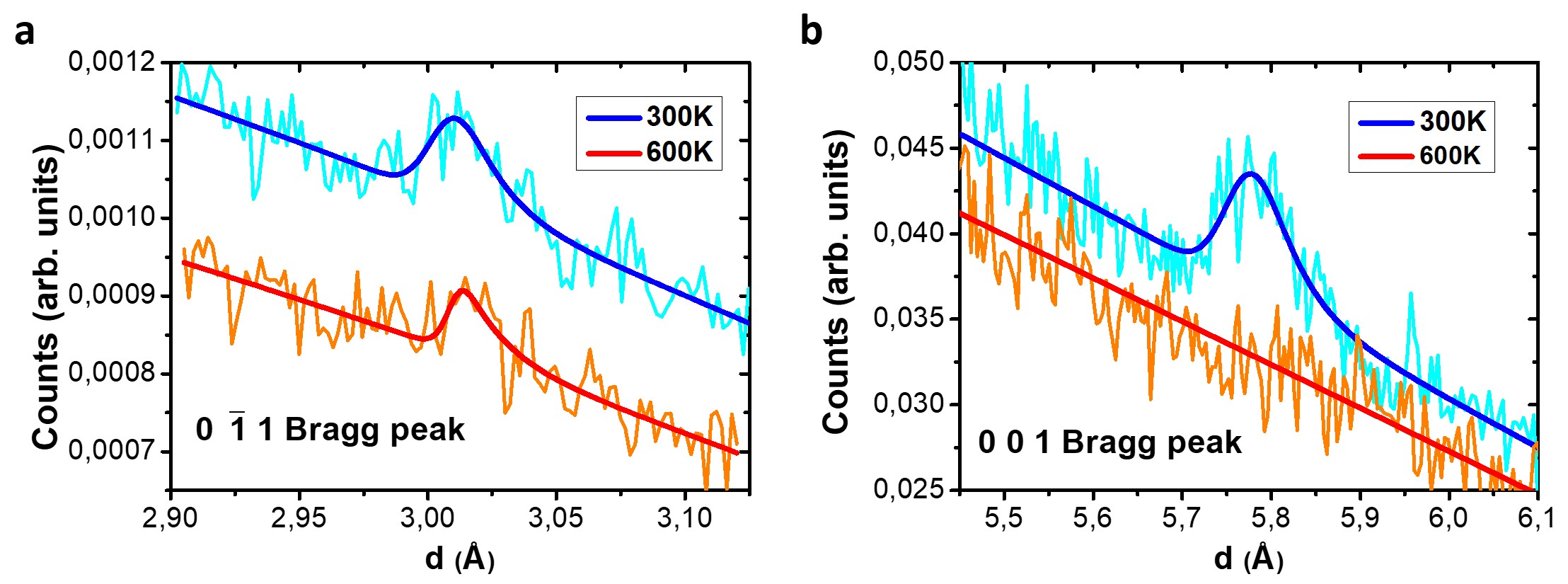} 
\caption{\label{neutron} 
{{\bf Neutron diffraction measurements of Cr$_{0.75}$Mn$_{0.25}$Sb (100) epitaxial thin films.} {\bf a} Structural and magnetic contributions to the $0\bar{1}1$ Bragg reflection measured below (300~K) and above (600~K) the N\'eel temperature. {\bf b} Purely magnetic 001 Bragg reflection at 300~K and its disappearance above the N\'eel temperature (600~K). Thin and thick lines denote the experimental data and corresponding fits, respectively.}
}
\end{figure*}

{\bf Neutron diffraction on Cr$_{0.75}$Mn$_{0.25}$Sb(100)}\\
We focus on the magnetic phase in which the compensated antiparallel moments are tilted away from the c axis, thereby enabling an anomalous Hall response in the (100)-oriented thin-film geometry (see micromagnetic modeling below). To verify that this phase corresponds to the epitaxial Cr$_{0.75}$Mn$_{0.25}$Sb (100) thin films, we investigated the magnetic order by neutron diffraction. The critical signature for magnetic order is the 001 Bragg reflection, which is structurally forbidden because the Cr$_{0.75}$Mn$_{0.25}$Sb unit cell contains two Cr(Mn) planes perpendicular to the c axis that are equivalent for nuclear scattering.  The anticipated magnetic order breaks this equivalence by imposing opposite spin orientations on alternating Cr(Mn) planes. This magnetic contribution enters the neutron-scattering cross section only if a component of the magnetization is perpendicular to the scattering vector (see {\it Experimental Methods}). Accordingly, the observation of the 001 reflection in the neutron-diffraction data shown in Fig.~\ref{neutron} confirms both the alternating magnetization of successive Cr(Mn) planes and the presence of a component of the magnetic moments in the hexagonal plane.

To quantify the magnetic structure, we fitted the neutron intensities of the $0\bar{1}1$ reflection, which contains both nuclear and magnetic contributions, together with those of the 001 reflection measured at room temperature and above the magnetic ordering temperature (600~K), using FullProf software \cite{Full, Rod93} (Fig.~\ref{neutron}). The fit yields a Cr(Mn) magnetic moment of $2.3(2)\mu_{\rm B}$ tilted by 35$^{\circ}$ with respect to the c axis (see {\it Experimental Methods)}.

As this magnetic structure is consistent with the occurrence of an anomalous Hall effect, we focus the remainder of this study on Cr$_{0.75}$Mn$_{0.25}$Sb(100).

\vspace{5pt}

\begin{figure}
\includegraphics[width=1.0\columnwidth]{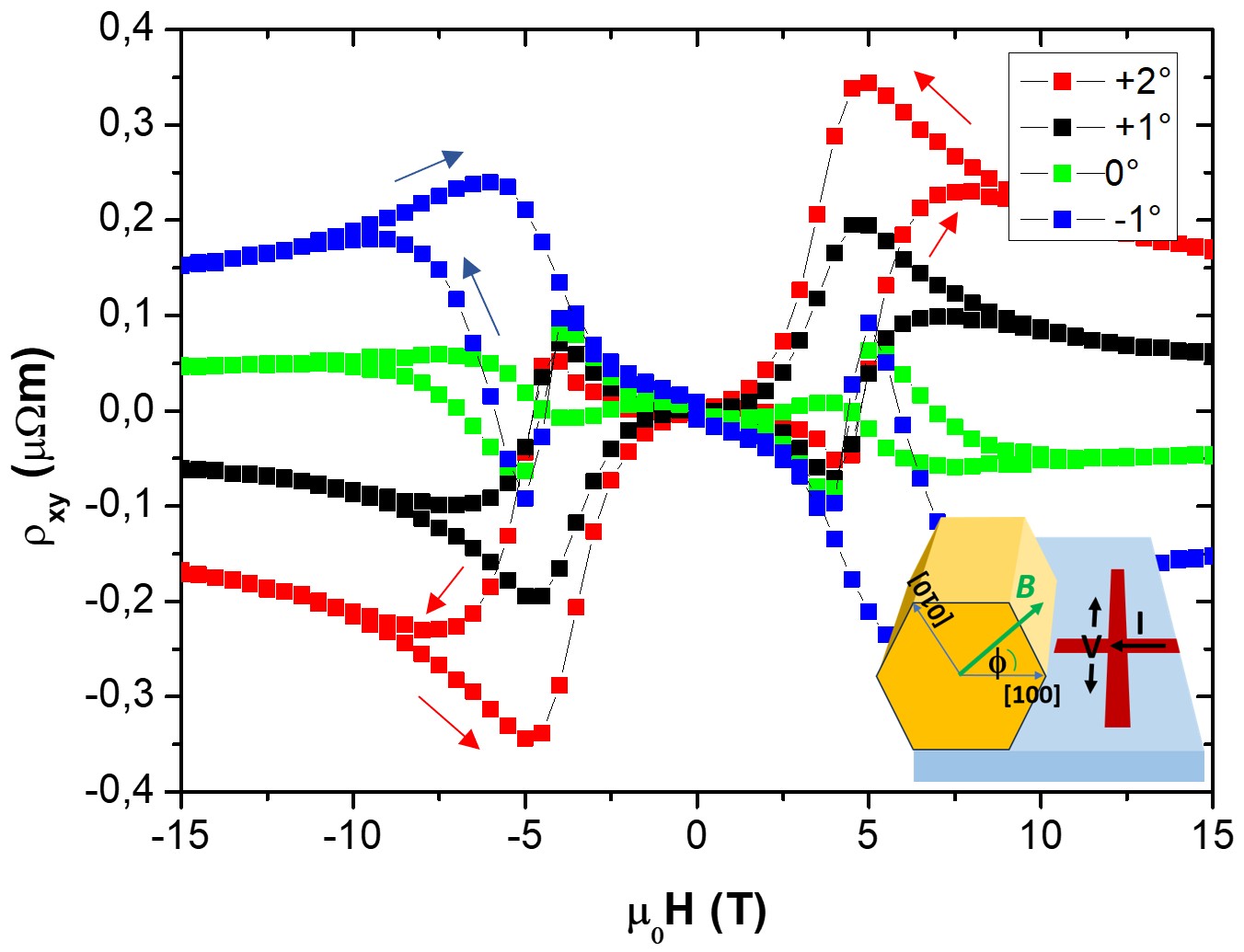} 
\caption{\label{Fig2} 
{\bf Hall-resistivity of Cr$_{0.75}$Mn$_{0.25}$Sb(100).} Magnetic field dependence of the Hall-resistivity at 300~K for field directions close to the in-plane [100]-direction. The inset shows a schematic representation of the sample geometry indicating the angle $\phi$ by which the magnetic field direction is tilted with respect to the in-plane [100]-direction. The yellow hexagonal prism indicates the orientation of the CrSb unit cell with respect to the substrate plane (blue).} 
\end{figure}

{\bf Anomalous Hall-effect in Cr$_{0.75}$Mn$_{0.25}$Sb(100)}\\
In epitaxial Cr$_{0.75}$Mn$_{0.25}$Sb (100) thin films at 300~K, we observe a strongly nonlinear Hall response featuring two hysteresis loops, which we attribute to the AHE. Here, we highlight field directions close to the in-plane [100]-direction, in which case the symmetry with respect to the film plane is broken by a small component of the magnetic field while contributions of the normal Hall effect are kept minimal. Figure~\ref{Fig2} shows a pronounced increase in the Hall voltage within the magnetic-field range of 3--5~T when the field direction is slightly tilted away from the in-plane [100] direction of the sample; the measurement geometry is illustrated in the inset.
Corresponding measurements covering a wide range of field tilt angles are shown in the Supplementary Material, section III.

Notably, none of the other Cr$_{1-x}$Mn$_{x}$Sb thin film compositions shown in Fig.\,\ref{Fig1} exhibit comparable behavior (see section III of the Supplementary Information), although a strain-induced AHE has recently been reported for epitaxial CrSb(100) thin films \cite{Zho25}.
Contributions from ferromagnetic impurities to our Hall signal or a field induced metamagnetic transition can be excluded on the basis of field dependent magnetization measurements (see section II of the Supplementary Information).

The hysteretic nonlinearity observed in the Hall effect of our Cr$_{0.75}$Mn$_{0.25}$Sb (100) thin films therefore indicates that a magnetic field $\mu_0\mathbf{H}$ in the range of 3--5~T, applied at a specific orientation, allows to align the N\'eel vector $\mathbf{n}$ into a configuration that enables the observation of an AHE. As we show below, the unusual Hall response originates from a transition involving eight distinct compensated magnetic configurations. In a magnetic field, these configurations separate into two groups with different energies. One group is energetically favored for $|B|<5$~T, whereas the other includes the ground state for $|B|>5$~T.

To characterize the magnetic anisotropy, which is essential for modeling the magnetic field driven motion of the N\'eel vector, we use angular dependent measurements of the longitudinal anisotropic magnetoresistance (AMR).

\vspace{5pt}
{\bf Longitudinal anisotropic magnetoresistance of Cr$_{0.75}$Mn$_{0.25}$Sb(100)}

\begin{figure}
\includegraphics[width=1.04\columnwidth]{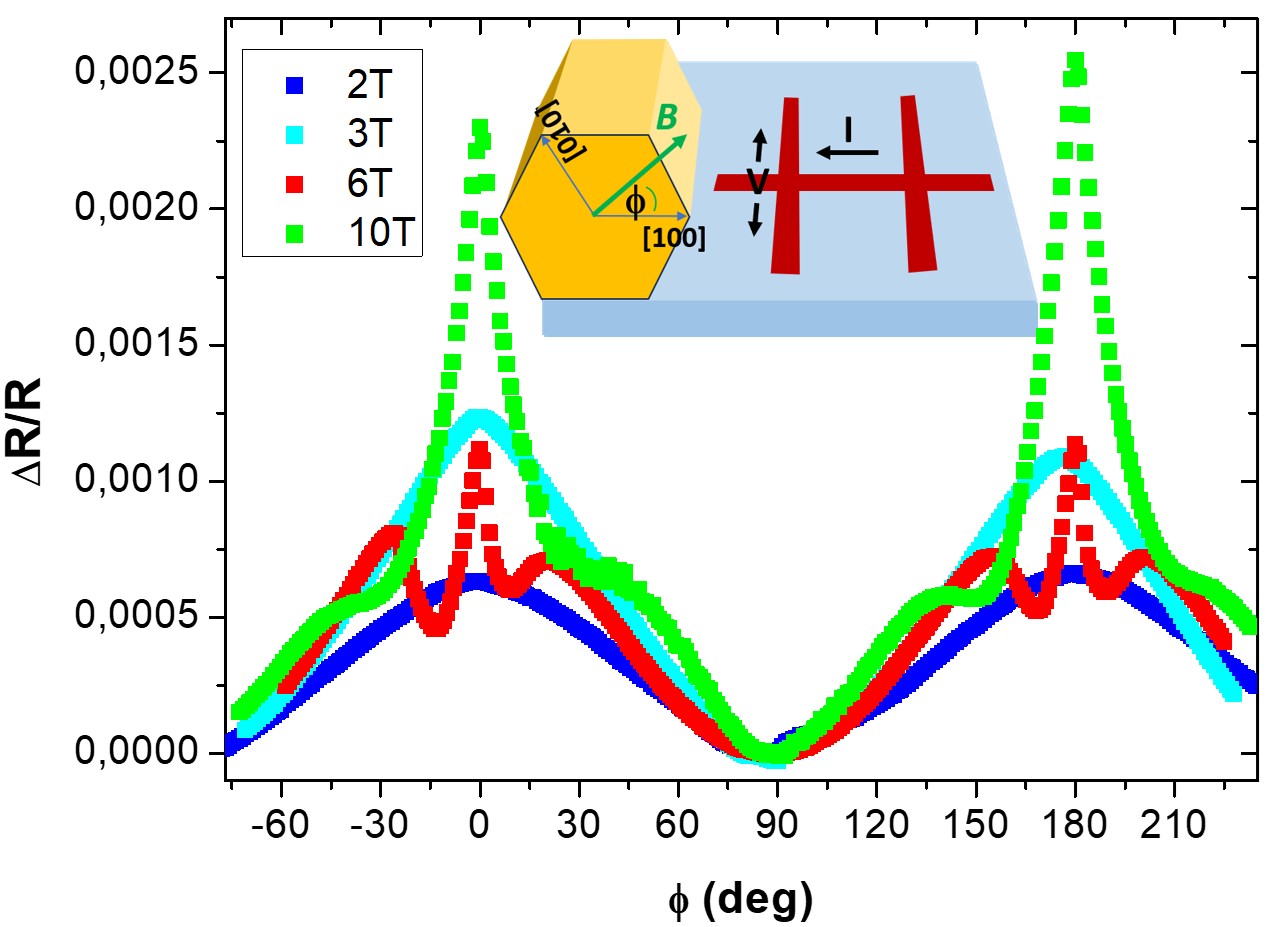} 
\caption{\label{Fig3} 
{\bf Longitudinal magnetoresistance (AMR) of Cr$_{0.75}$Mn$_{0.25}$Sb(100).} Dependence of the longitudinal resistance on the orientation of magnetic fields with different magnitude (at 300~K). The inset shows a schematic representation of the sample geometry indicating the angle $\phi$ by which the magnetic field direction is tilted with respect to the in-plane [100]-direction.} 
\end{figure}

Within the hexagonal c-plane, the magnetocrystalline anisotropy is generally weak so that even moderately strong magnetic fields can reorient the c-plane component of the N\'eel vector. Assuming perfect hexagonal symmetry, a magnetic field rotating by an angle $\phi$ around the c-axis of Cr$_{0.75}$Mn$_{0.25}$Sb(100) would therefore reflect this weak six-fold symmetry and induce a corresponding smooth rotation of the N\'eel vector. However, we observe a two-fold rotational symmetry of the longitudinal resistance R$_{\rm AMR}(\phi)$, as shown in Fig.~\ref{Fig3}. This symmetry can be explained by uniaxial strain within the film plane, which is originating from the epitaxial growth on GaAs(110), i.\,e.\,, on a substrate cut with a two-fold in-plane symmetry.

At sufficiently large magnetic fields ($>5$\,T), the angular dependence of the AMR R$_{\rm AMR}(\phi)$, exhibits a pronounced and sharp peak at $\phi=0^{\rm o}$. This peak indicates that, within the c-plane, the [100] direction is energetically favored, allowing to infer that, in zero field, the N\'eel vector lies within the film plane. When a sufficiently strong magnetic field is applied along the [100] direction, it induces a rotation of the N\'eel vector into a state with an out-of-plane component--analogous to a spin-flop transition--thereby modifying the AMR.

With this insight, we are able to model the field-driven reorientation of the N\'eel vector and its impact on the Hall response, which is essential to unambiguously attribute the observed nonlinear Hall signal to an intrinsic AHE arising from the altermagnetic nature of the system.

\vspace{5pt}
{\bf Modeling N\'eel vector orientation, AHE and AMR of Cr$_{0.75}$Mn$_{0.25}$Sb(100)}

\begin{figure}
\includegraphics[width=1.0\columnwidth]{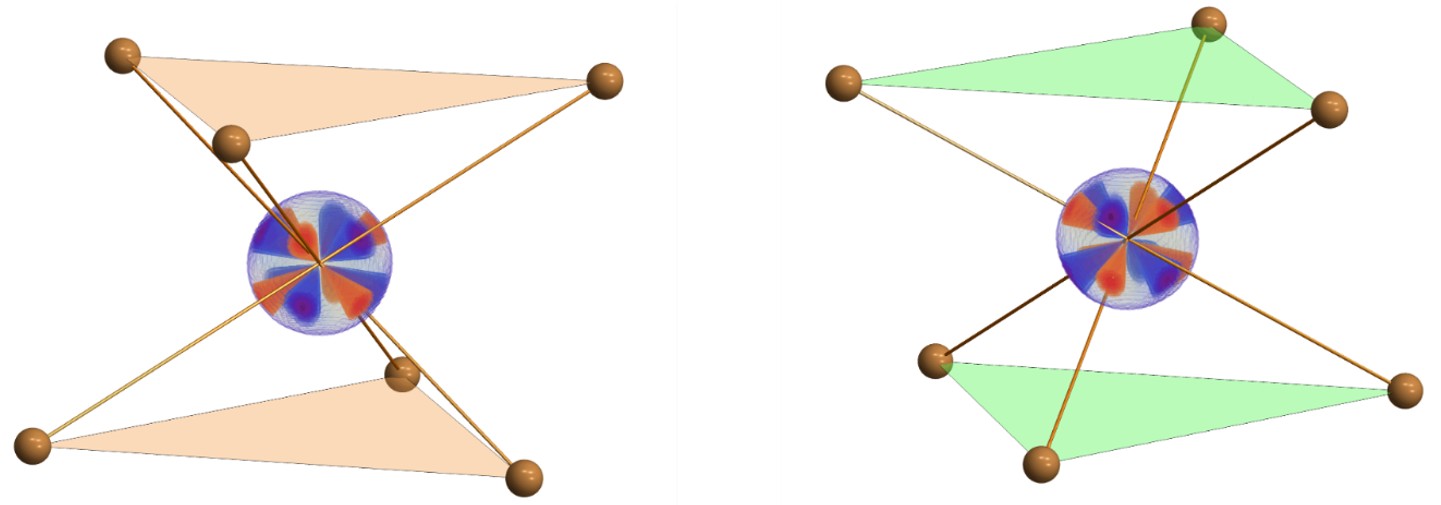}
\caption{\label{Fig4} 
{{\bf Altermagnetic order parameter.} Schematic representation of the two altermagnetic sublattices with the Cr atom in the center and the surrounding Sb atoms. The red and blue lobes represent the spherical harmonics whose difference represents the altermagntic order parameter.}} 
\end{figure}

Our micromagnetic modeling of the dynamics of the N\'eel vector ${\bf n=M_1-M_2}$ and the total magnetization ${\bf m=M_1+M_2}$ is guided by Landau’s theory of magnetic phase transitions. Within this framework, the free energy is expressed as a power-series expansion in a generalized order parameter, constrained by the symmetries of the system.
For a conventional antiferromagnet, this order parameter is simply the N\'eel vector. In contrast, collinear altermagnetism arises from the symmetry of the non-magnetic atomic environment surrounding the magnetic sublattices, such that mapping one sublattice onto the other requires a combined partial translation within the unit cell and a rotation. Consequently, an adequate altermagnetic order parameter must be used to capture this local symmetry \cite{Gom24}.
We employ spherical harmonics to represent the orientation of the electronic bonds associated with each magnetic sublattice. The red and blue lobes shown in Fig.\,\ref{Fig4} indicate positive and negative weights of these bonds, respectively, thereby defining an anisotropy profile for the N\'eel vector $\bf n$ and the magnetization ${\bf m}$. The altermagnetic order parameter $Q_{\rm AM}$ is then defined as the difference between the corresponding spherical harmonics (see Supplementary Information).

We derive the equilibrium orientations of $\bf n$ and $\bf m$  by minimizing the free-energy functional of Cr$_{1-x}$Mn$_x$Sb in the presence of an external magnetic field $\bf{H}$. As detailed in the Supplementary Information, the free energy includes an exchange term that enforces antiparallel alignment of the sublattice magnetizations $\bf{M_1}$ and $\bf{M_2}$, as well as a Zeeman term that aligns $\bf m$ parallel to $\bf{H}$. In addition, Dzyaloshinskii–Moriya interactions (DMI) arise from a combination of the exchange and local anisotropy fields of the parent hexagonal lattice, and from symmetry breaking induced by growth-related strain within the film plane.

\begin{figure}
\includegraphics[width=1.0\columnwidth]{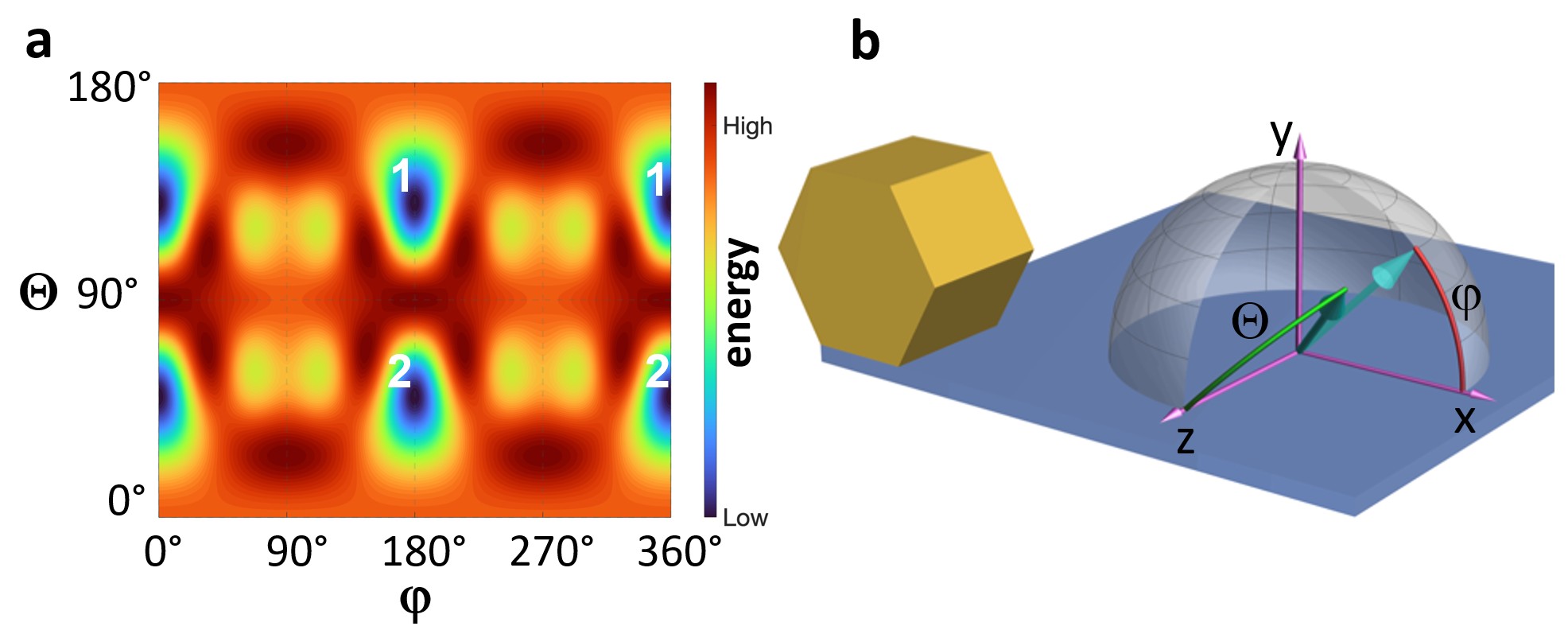}
\caption{\label{Fig5} 
{{\bf N\'eel vector orientation and free energy.} {\bf a} Color scale representation of the dependence of the free energy on the N\'eel vector orientation. $\theta$ represents the polar angle with respect to the z-direction and $\varphi$ the azimuthal angle of the N\'eel vector. {\bf b} Corresponding coordinate system illustrating the orientation of the unit cell relative to the substrate.}} 
\end{figure}

By keeping the number of free parameters to a minimum, we are able to reproduce the experimentally observed tilting of the N\'eel vector away from the hexagonal c-axis by considering a single DMI term imposed by the hexagonal symmetry, resulting in a twelvefold degeneracy of the ground state N\'eel-vector orientations. The inclusion of an additional in-plane strain–induced DMI term partially lifts this degeneracy, resulting in four remaining energetically equivalent directions as shown in the energy  map in Fig.\,\ref{Fig5}. The calculated free-energy landscape exhibits four absolute minima (dark blue in the color map), corresponding to four equivalent ground-state orientations of the N\'eel vector. Among these, those labeled {\it 1} are associated with a DMI induced weak magnetization parallel to the $y$-direction, whereas those labeled {\it 2} correspond to an antiparallel orientation.

We further use symmetry considerations to model the relationship between the AHE and the N\'eel vector.  As described in detail in the Supplementary Information, we expand the general resistivity tensor $R_{ij}$ in terms of the magnetic variables and retain only the lowest order terms consistent with the point symmetry group of the nonmagnetic phase. Due to strong exchange coupling between sublattices, we note that the value of magnetization is negligibly small. However, the magnetization vector has the same symmetry as the Hall vector and can therefore be used to describe the field dependence of the resistivity. In our measurement geometry with the current applied along the $x$-direction, the AHE is therefore sensitive to the $z$-component of the N\'eel vector, $n_z$, and to the $y$-component, $\hat{m}_y$,  of the  unit vector $\hat{\mathbf{m}}$ parallel to magnetization :
\begin{equation}\label{eq_AHE}
    \rho_\mathrm{AHE}=\alpha_\mathrm{AHE}Q_\mathrm{AM} n_z+\beta_\mathrm{AHE} \hat{m}_y(\mathbf{n}),
\end{equation}
where $Q_{\rm AM}$ is the altermagnetic order parameter, $ \alpha_\mathrm{AHE}$ and $\beta_\mathrm{AHE}$ are phenomenological constants.

All four ground state orientations of the N\'eel vector shown in Fig.\,\ref{Fig5} are energetically degenerate and therefore occur with equal probability, resulting in an equal population of domains in zero magnetic field. Averaging over these domains naturally explains the absence of an AHE in the experimental data at zero field.

\begin{figure}
\includegraphics[width= 0.9\columnwidth]{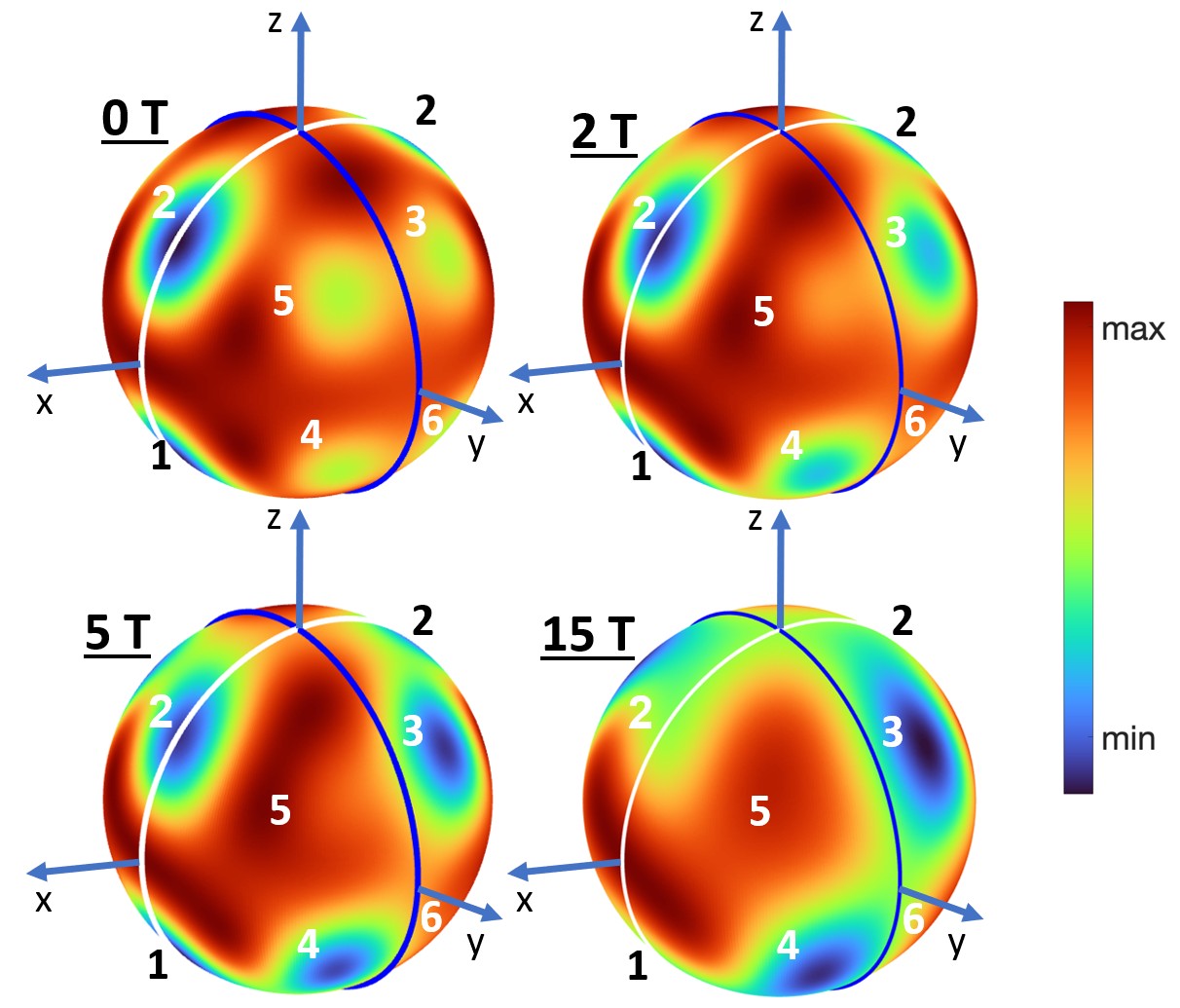}
\caption{\label{Fig6} 
{{\bf Field driven N\'eel vector orientation.} Energy maps associated with the N\'eel vector orientation projected on a sphere for different magnitudes of the magnetic field tilted by $\phi=2^{\rm o}$ from the x-direction towards the y-direction.}} 
\end{figure}

However, we experimentally observed an AHE when the magnetic field is tilted slightly out of the film plane (Fig.\,\ref{Fig2}).
Accordingly, in our simulations we consider the application of a magnetic field tilted by $\phi=2^{\rm o}$ from the $x$-axis (in-plane [100]-direction) towards the $y$-axis. The corresponding energy landscapes, shown in Fig.\,\ref{Fig6} as a projection onto a sphere, illustrate how the energy minima evolve with increasing magnetic field. 

A transition of the absolute energy minima from positions {\it 1} and {\it 2} at zero field to position {\it 3} occurs at a magnetic field of $\simeq 5$~T. This field-induced reorientation of the N\'eel vector is therefore suggested as the source of the pronounced non-linearity and hysteresis observed experimentally in the Hall-effect within this field range.

\begin{figure*}
\includegraphics[width= 1.8\columnwidth]{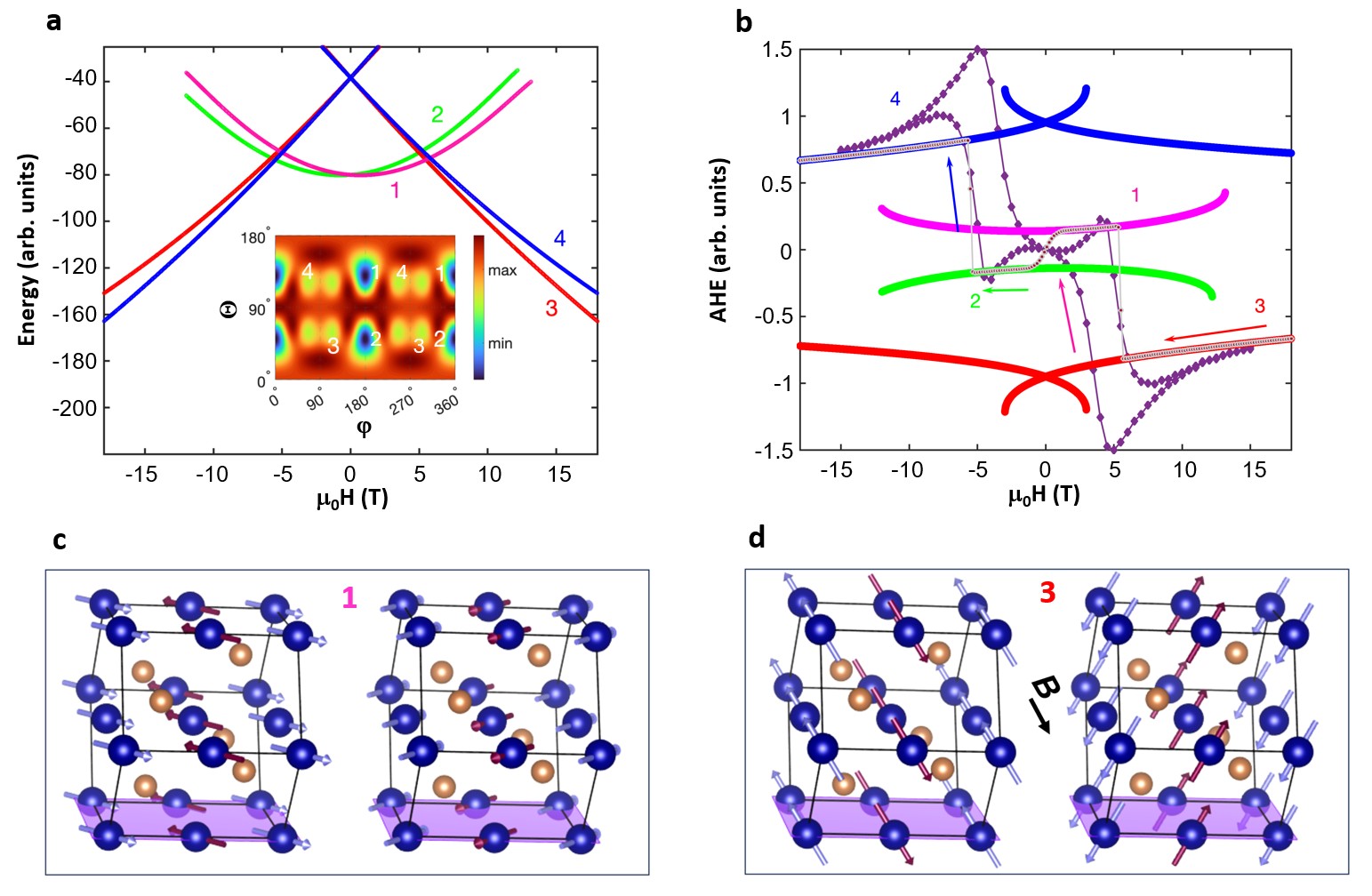}
\caption{\label{Fig7} 
{{\bf Field driven N\'eel vector orientation and AHE.} Magnetic field tilted by $\phi=2^{\rm o}$ from the x-direction towards the y-direction. Panel {\bf a} shows the energy associated with the N\'eel vector orientations labeled in the inset. Panel {\bf b} shows the AHE corresponding to these orientations. By sweeping from positive to negative magnetic fields, the lowest energy orientation of the N\'eel vector changes from state 3 to state 1 to state 2 to state 4 corresponding to the AHE indicated by the gray data points. The violet data point show experimental results for comparison. Panel {\bf c} shows the orientation of the magnetic moments corresponding to both state 1 configurations and panel {\bf d} those corresponding to state 3. Corresponding images for the states 2 and 4 are shown in the Supplementary Information, Fig. S12.}} 
\end{figure*}

Panel {\bf a} of Fig.\,\ref{Fig7} displays the energy of all local minima that attain the lowest overall energy within any part of the experimentally accessed field range. Minima that are energetically equivalent in our geometry are assigned the same label in the inset showing the zero field energy map.

Using Eq.\eqref{eq_AHE}, we calculate the AHE associated with these energy minima. Panel {\bf b} of Fig.\,\ref{Fig7} presents the AHE corresponding to the N\'eel vector orientations labeled {\it 1-4}. The contributions of both terms in Eq.\eqref{eq_AHE} are found to be comparable in magnitude and are therefore shown separately in the Supplementary Information.

Next, we consider the field-dependent evolution of the N\'eel vector and the corresponding AHE in analogy to our experiments. We consider starting a magnetic field sweep from a large positive value. In this regime, the N\'eel vector orientation labeled {\it 3} is lowest in energy (red line in panel {\bf a} of Fig.\,\ref{Fig7}), and the corresponding AHE (red line in panel {\bf b}) is negative, in agreement with the experimental data (violet data points). At $\simeq 5$~T, N\'eel vector configuration {\it 1} becomes energetically favorable, resulting in a relatively small positive AHE (pink curve). The corresponding magnetic structures are illustrated in panels {\bf c} and {\bf d}. The predicted sign reversal of the AHE is in good agreement with the experimental data, which additionally exhibit hysteresis, as expected for a multidomain state in which domain growth is governed by domain-wall motion and pinning.
At zero magnetic field, states {\it 1} and {\it 2} are degenerate. Consequently, the corresponding multidomain state exhibits a vanishing AHE.
Upon further decreasing the magnetic field to negative field values, configuration {\it 2} becomes energetically favorable, leading to another sign change of the AHE (green curve). Finally, at $\simeq 5$~T, configuration {\it 4} corresponds to the energy minimum, resulting in a relatively large positive AHE (blue curve). 

Overall, the calculated AHE corresponding to the lowest energy orientation of the N\'eel vector (gray curve in Fig.\,\ref{Fig7}{\bf b}) shows good qualitative agreement with the experimentally observed Hall-effect for magnetic fields tilted by $2^{\rm o}$ out of the sample plane from the [100]-direction (pink data points in Fig.\ref{Fig7}{\bf b}). This agreement strongly supports the conclusion that the hysteretic nonlinearities observed experimentally in the Hall effect of Cr$_{0.75}$Mn$_{0.25}$Sb(100) originate from an AHE enabled by altermagnetism.

In summary, partial substitution of Cr by Mn in epitaxial CrSb(100) thin films modifies the magnetic  easy direction, thereby engineering the symmetry that enables the occurrence of an anomalous Hall effect (AHE). Guided by Landau theory, we model the field driven orientation of the N\'eel vector and the associated AHE, achieving good qualitative agreement with the experimental observations. This allows us to reliably conclude that the pronounced hysteretic and nonlinear behavior observed in the magnetic field dependence of the Hall signal is indeed an AHE arising from the altermagnetic nature of Cr$_{0.75}$Mn$_{0.25}$Sb. Our experiments demonstrate the feasibility of tuning the electronic and magnetic properties of established altermagnets to meet the required symmetry for electrical read-out in spintronics applications. In particular, we have shown that CrSb, an altermagnet with exceptionally high ordering temperature and electrical conductivity, can be engineered to enable electrical read-out of the N\'eel vector orientation via measurements of the AHE. 

\section{Experimental Methods}

{\bf Sample preparation and characterization}

Epitaxial Cr$_{1-x}$Mn$_{x}$Sb(100) thin films with a thickness of 30~nm were deposited on GaAs(110) substrates by DC magnetron sputtering from a single cathode using a composite target composed of elemental segments of the constituent materials Cr, Mn, and Sb, each occupying either 1/16 or 1/8 of the total target area. Varying fractions of the Cr segments were replaced with Mn to obtain different magnetic phases of Cr$_{1-x}$Mn$_{x}$Sb. After deposition at a substrate temperature of $\simeq300^{\rm o}$C, the samples were annealed in-situ at a temperature of $\simeq400^{\rm o}$C for 15~min. Thin films with a thickness of 40~nm were deposited with a rate of 2~nm/min. These deposition parameters were identical to those used for thin films of the parent compound CrSb \cite{Rei24}.

Crystallographic characterization of the samples, presented in the Supplementary Information, was carried out by X-ray diffraction using a Bruker D8 Discover four-circle diffractometer and by scanning transmission electron microscopy (STEM) using a FEI Titan TEM.
\\

{\bf Magnetotransport measurements}

Magnetotransport measurements (Figs.\,2 and 4) in magnetic fields below 15~T were performed using an Oxford Instruments Cryostat equipped with a variable temperature inset, a superconducting magnet, and a rotatable sample stage. The measurements were conducted with a Keithley 6220 precision current source supplying a probe current of 10~$\mu$A and a Keithley 2182A Nanovoltmeter operated in Delta mode, with averaging over 100 measurements.
The AMR measurement in magnetic fields above 15~T (Fig.\,1{\bf d}) were carried out at the Dresden High Magnetic Field Laboratory using pulsed magnetic fields with a duration of 100~ms. Data were acquired at a sampling rate of 80~kHz using a numerical lock-in technique \cite{Bod20}.

The samples were lithographically patterned into stripes and crosses aligned with the crystallographic directions of the thin films as indicated in Figs.\,1, 2, and 3.
\\

{\bf Magnetization measurements}

Magnetization measurements (see Supplementary Information) were performed using a superconducting quantum interference device (SQUID) magnetometer (Quantum Design MPMS), equipped with a 5~T superconducting magnet. 
\\

{\bf Neutron diffraction}

Neutron diffraction experiments on an epitaxial Cr$_{0.75}$Mn$_{0.25}$Sb(100) thin film (100~nm thick) were performed on the time-of-flight diffractometer WISH at the ISIS Neutron and Muon Source \cite{Cha11}. The film was aligned in the (0~K~L) scattering plane, and the 0${\bar 1}$1 and 001 reflections were optimized with respect to the incident neutron flux. Details of the geometry can be found in reference \cite{Orl26}. Owing to the small sample volume, an accumulation time of 17 h per scan was required for appropriate signal-to-noise ratio.
To obtain a paramagnetic reference, the film was measured at 600~K. At this temperature, no intensity was observed at the 001 reflection, while the intensity of the 0${\bar 1}$1 reflection was significantly reduced. The 600~K data were used to determine the nuclear scale factor and to isolate the magnetic contribution to the 0${\bar 1}$1 reflection measured at room temperature. The magnetic intensities of the 0${\bar 1}$1 and 001 reflections were subsequently used to refine both the magnetic moment magnitude and the orientation of the N\'eel vector within the film plane. As mentioned above, the best fit yields a Cr(Mn) magnetic moment of $2.3(2)\mu_{\rm B}$, tilted by 35$^{\circ}$ with respect to the c-axis. A comparably good refinement of the room temperature data is obtained for a model with a substantially smaller magnetic moment of $1.1(2)\mu_{\rm B}$ and the N\'eel vector aligned along the in-plane a-axis. However, we consider the latter fit solution less likely, because the resulting moment is significantly smaller than expected based on measurements of bulk samples \cite{Rei82} and the obtained scale factor significantly overestimates the intensity of the 0${\bar 1}$1 reflection measured at 600~K. 
\\

{\bf Acknowledgements}

We acknowledge funding by the Deutsche Forschungsgemeinschaft (DFG, German Research Foundation) - TRR 173 - 268565370 (projects A05, A11, A01, and B02)  and JO 404/12-1. We acknowledge support by Fabio Orlandi at WISH.
\\

{\bf Author contributions}

M.G.F., L.O., O.G. and M.J. wrote the paper with contributions by S.R. and D.K.; M.G.F. and  L.O. prepared the samples; M.G.F., L.O., and M.J. performed all transport measurements; the neutron diffraction experiments were carried out by P.M., D.K., S.L., L.O. and M.J. with D.K. refining the results; O.G. simulated the AHE based on her theoretical model; T.D., J.V.V, and R.D.-B. provided the STEM images; M.K. and J.S. contributed to the discussion of the results and provided input; M.J. initiated and coordinated the project. 
\\

{\bf Competing interests:}

 The authors declare no competing financial interests.

\end{document}